%% file: bare_jrnl.tex
\documentclass[journal]{IEEEtran}
\usepackage{lipsum}

\usepackage{cite}

\usepackage{csquotes}

\usepackage{svg}
\usepackage{mwe}
\usepackage{tikz}
\usepackage{tikz-3dplot}
\usepackage{pgfplots}
\pgfplotsset{compat=1.6}
\usetikzlibrary{arrows, decorations.markings}
\usetikzlibrary{calc}
\usetikzlibrary{shapes}
\usetikzlibrary{arrows,decorations.pathmorphing,backgrounds,positioning,fit,petri}

\definecolor{review}{rgb}{0,1,0}
\definecolor{change}{rgb}{1,0,0}
\definecolor{revised}{rgb}{0,0,0}

\definecolor{max}{named}{orange}

\usepackage[nooldvoltagedirection]{circuitikz}
\usepackage{overpic}
\ifCLASSINFOpdf
\else
\fi
\usepackage{todonotes}

\usepackage{amsmath}
\usepackage{siunitx}
\DeclareSIUnit{\fps}{fps}

\usepackage{booktabs}
\begin{document}
%

\title{Infectious Disease Transmission via Aerosol Propagation from a Molecular Communication Perspective: Shannon Meets Coronavirus}
%
%
%

\author{Max~Schurwanz,
        Peter Adam~Hoeher, 
        Sunasheer~Bhattacharjee,
        Martin~Damrath,
        Lukas~Stratmann, 
        and~Falko~Dressler, 
\thanks{M.~Schurwanz, P.\,A.~Hoeher, S.~Bhattacharjee, and M.~Damrath are with the Faculty of Engineering, Kiel University, Kiel, Germany, e-mail: \{masc,ph,sub,md\}@tf.uni-kiel.de.}
\thanks{L.~Stratmann and F.~Dressler are with the School of Electrical Engineering and Computer Science, TU Berlin, Berlin, Germany, e-mail: \{stratmann,dressler\}@ccs-labs.org.}
}

\maketitle

\input{0_abstract}

%
\IEEEpeerreviewmaketitle

\input{0_keywords}
\input{1_introduction}
\input{2_scenario}
\input{3_minimization}
\input{4_simulation}
\input{5_open_tasks}
\input{6_conclusion}
\input{7_appendix}
\input{8_acknowledgement}

\ifCLASSOPTIONcaptionsoff
  \newpage
\fi



\bibliographystyle{IEEEtran}
\bibliography{IEEEabrv,literature}
%



%


\vspace{-2ex}
\begin{IEEEbiographynophoto}{Max Schurwanz}
received the M.Sc. degree in electrical engineering from Kiel University, Germany, in 2020, where he is currently pursuing the Dr.-Ing. (Ph.D.) degree at the Faculty of Engineering. Since 2020, he is a Research Assistant at the Chair of Information and Coding Theory, Kiel University. His current research interests include air-based macroscopic molecular communication and radar signal processing.
\end{IEEEbiographynophoto}
\vspace{-2em}
\begin{IEEEbiographynophoto}{Peter Adam Hoeher}
 received the Dipl.-Ing. (M.Sc.) degree in electrical engineering from RWTH Aachen University, Aachen, Germany, in 1986, and the Dr.-Ing. (Ph.D.) degree in electrical engineering from the University of Kaiserslautern, Kaiserslautern, Germany, in 1990. From 1986 to 1998, he was with the German Aerospace Center (DLR), Oberpfaffenhofen, Germany. From 1991 to 1992, he was on leave at AT\&T Bell Laboratories, Murray Hill, NJ. Since 1998 he is a Full Professor of electrical and information engineering at Kiel University, Kiel, Germany. His research interests are in the general area of wireless communications and applied information theory. 
\end{IEEEbiographynophoto}
\vspace{-2em}
\begin{IEEEbiographynophoto}{Sunasheer Bhattacharjee}
received the M.Sc. degree in digital communications from Kiel University, Germany, in 2018, where he is currently pursuing the Dr.-Ing. (Ph.D.) degree at the Faculty of Engineering. Since 2019, he has
been a Research and Teaching Assistant at the Chair of Information and Coding Theory, Kiel University. His current research interests include physical layer issues of molecular communication and testbed design.
\end{IEEEbiographynophoto}
\vspace{-2em}
\begin{IEEEbiographynophoto}{Martin Damrath}
 received the M.Sc. degree in electrical engineering from Kiel University, Germany, in 2014. Since 2014, he has been a Research and Teaching Assistant at the Chair of Information and Coding Theory, Kiel University. In October 2020, he defended his Dr.-Ing. (Ph.D.) thesis at the Faculty of Engineering. His current research interests include physical layer issues in the area of molecular communications, including high-order modulation schemes, channel coding, advanced detection, and multiple access. 
\end{IEEEbiographynophoto}
\vspace{-2em}
\begin{IEEEbiographynophoto}{Lukas Stratmann}
is pursuing his Ph.D. degree with focus on the simulation of macroscopic molecular communication.
He began his Ph.D. studies at Paderborn University in 2019, where he also completed his M.Sc.\ degree with a thesis on visual attention in a virtual cycling environment in 2019 and his Bachelor degree in 2017.
In 2020, he joined his colleagues in merging with the Data Communications and Networking group at TU Berlin.
\end{IEEEbiographynophoto}
\vspace{-2em}
\begin{IEEEbiographynophoto}{Falko Dressler}
 is full professor and Chair for Data Communications and Networking at the School of Electrical Engineering and Computer Science, TU Berlin. He received his M.Sc. and Ph.D. degrees from the Dept. of Computer Science, University of Erlangen in 1998 and 2003, respectively.
His research objectives include adaptive wireless networking (sub-6GHz, mmWave, visible light, molecular communication) and wireless-based sensing with applications in ad hoc and sensor networks, the Internet of Things, and cyber-physical systems.
\end{IEEEbiographynophoto}







\end{document}

%% file: 0_abstract.tex
\begin{abstract}
Molecular communication is not only able to mimic biological and chemical communication mechanisms, but also provides a theoretical framework for viral infection processes.
In this tutorial, aerosol and droplet transmission is modeled as a multiuser scenario with mobile nodes, related to broadcasting and relaying. In contrast to data communication systems, in the application of pathogen-laden aerosol transmission, mutual information between nodes should be minimized. Towards this goal, several countermeasures are reasoned. The findings are supported by experimental results and by an advanced particle simulation tool. This work is inspired by the recent outbreak of the coronavirus (COVID-19) pandemic, but also applicable to other airborne infectious diseases like influenza.
\end{abstract}

%% file: 0_keywords.tex

%% file: 1_introduction.tex
\section{Introduction}\label{sec:introduction}
\IEEEPARstart{M}{olecular} communication (MC) aims at providing technical solutions to real-world problems using techniques borrowed from nature, that have been adapted over the course of evolution~\cite{Nakano2013}. Even in the very beginning of MC research, networking foundations have been considered~\cite{Akyildiz2008}. Communication links can be established in situations, where traditional electromagnetic communication techniques pose a challenge. In contrast to classical communication paradigms, molecules are used as information carriers, enabling biocompatible solutions. In the past decade, various MC applications have been identified, both in the microscopic and the macroscopic domain~\cite{farsad2016survey}. Examples include sensing of target substances in biotechnology, smart drug delivery in medicine, and monitoring of oil pipelines or chemical reactors in industrial settings~\cite{Jamali2019}.

The global pandemic of the \textit{severe acute respiratory syndrome coronavirus 2} (SARS-CoV-2) inspires researchers from numerous disciplines to combine different areas to find new and innovative solutions to urgent problems. The need for novel solutions has already created rapid infection control solutions during the pandemic. These include ``social distancing'' rules~\cite{Glass2006}, mouth and nose protection in public spaces, and large-scale disinfection measures as suggested by the various institutes of health care and virology around the world. {\color{revised} Since MC combines biological/chemical and communication/ information theoretical approaches, a common framework can be established between MC and infectiology} to apply known techniques from information engineering to the field of health care and infection prevention. Findings from health research dealing with the spread of aerosols and the viruses they may contain, can be joined with insights from MC in order to predict infection processes and to identify more sophisticated and safer protective measures.

Recent research in combining these areas has dealt with possible use cases for aerosol communications~\cite{Khalid2019}, and with channel modeling considering infectious aerosols in point-to-point scenarios~\cite{Khalid2020,Gulec2020}. An infected person acts as a transmitter of pathogen-laden aerosols and a device or another person acts as a receiver of the infectious particles. The close relation to macro-scale MC, where a transmitter emits molecules into a fluid environment and an absorbing receiver recovers the information, is exploited. In~\cite{Schurwanz2020}, the concept has been expanded to include several features:
\begin{itemize}
    \item The transmission of infectious aerosols is suggested to be a multiuser MC scenario.
    \item In each MC transmitter, the modulator is modeled by respiratory-event-driven higher order molecular variable-concentration shift keying~{\color{revised}(MoVCSK)}~\cite{Bhattacharjee2019}. 
    \item Channel modeling is aided by air-based human experiments using a fluorescence dye in conjunction with optical detection, motivated by the air-based macro-scale MC testbed published in~\cite{Bhattacharjee2020a}. 
    \item An advanced computer simulation tool is designed based on the work in~\cite{drees2020efficient}. Parameters feeding this tool were obtained by own experiments, among other established sources.
\end{itemize}

{\color{revised} The missing link between macro-scale MC and infection prevention} can be closed by the fields of multiuser communication and dynamic scenario modeling. Only when multiple users exchange particles in a dynamically changing way, reproduction -- the root of the pandemic -- can occur. Reproduction involves infection chains and coronavirus clusters. A large variety of safety measures can be rethought from a communication and information theory perspective to optimize the goal of infection prevention. 

Building on the novel contributions in~\cite{Schurwanz2020}, this tutorial presents the following communication aspects in greater depth: 
\begin{itemize}
    \item The transmission of infectious particles is modeled as a multiuser MC scenario with mobile nodes in a time-varying environment. Infected users are transmitters that broadcast pathogen-laden particles, while the remaining users are potential receivers. After {\color{revised} infection plus incubation times}, some receivers may become  active transmitters, like in relaying. {\color{revised} During this expected long delay, the environmental setup is likely to change completely.} 
    \item Infection is considered to be a generalized threshold detection problem: if the density of infectious particles received exceeds a user-dependent threshold, an infection is likely to occur. Furthermore, the viral load may also affect the course of the disease. In the field of MC, the corresponding analogy is reliability information.
    \item The viral infection rate is modeled by the concept of mutual information~{\color{revised} \cite{Akyildiz2019information,Hoeher2021}}.  Contrary to data transmission, the aim is to minimize the mutual information between an infected user and the healthy ones.
\end{itemize}

The remainder of this tutorial paper is structured as follows: Section~\ref{sec:scenario} describes the duality between {\color{revised} MC (comprising communication and information theory)} on the one hand and the field of airborne aerosol infection on the other.
In Section~\ref{sec:minimization_mutual_info}, the concept of mutual information is introduced to the area of infectious disease transmission.  
Prevention techniques from different areas are discussed to minimize the mutual information in order to reduce the risk of infection. 
Section~\ref{sec:simulation} presents a particle simulation tool suitable for the problem under investigation, and corresponding simulation results for a multiuser scenario.
Section~\ref{sec:open_questions} discusses the {\color{revised} prospects and challenges} that arise from the duality concept.  Finally, conclusions are drawn in Section~\ref{sec:conclusion}.

%% file: 2_scenario.tex
\section{Duality Aspects\label{sec:scenario}}
In the area of wireless communications, a general multiuser scenario consists of mobile nodes spatially distributed in a three dimensional (3D) environment, acting as transmitters and/or receivers. The communication links between all users are affected by environmental parameters, signal powers, and random distortions. 
{\color{revised} In the scenario under consideration, the channel characteristics are determined by the environment and the statistics of the aerosols and droplets emitted by the different users. The term aerosols is used in the literature for various particles sizes between approximately \SI{1}{\micro\meter} and several \SI{1000}{\micro\meter}. In this paper, aerosols describe smaller particles that experience negligible influence from gravitational forces whereas droplets describe larger particles that are forced to the ground relatively quickly by gravitational forces.}
{\color{revised} Analogous to the understanding in infectiology, it is assumed in this multiuser scenario that at least one transmitter emits infectious aerosols and droplets.} This situation can be understood as a point-to-multipoint communication scenario at the physical layer, related to broadcasting and relaying.  In case of a successful transmission -- a necessary condition is that the ingested amount of infectious aerosols exceeds the infection threshold  -- the receiver itself becomes an active transmitter after {\color{revised} infection plus} incubation period. 
{\color{revised} The infection time can be on the order of minutes, but the time required for a human to become an active transmitter can be up to several days.}
The {\color{revised} agile} relays act as virus spreaders or even superspreaders, i.e., they boost the viral load.

Table~\ref{table:duality} features some duality aspects between air-based MC and infectious particle transmission.
%
\begin{table*}[ht]
\addtolength\abovecaptionskip{-5pt}  
\addtolength\belowcaptionskip{0pt}  
\caption{Selected duality aspects.}
\centering
\begin{tabular}{lll}
\toprule
Device              & Air-based Molecular Communication     & Infectious Particle Transmission \\     
\midrule
Transmitter         & Sprayer                               & Mouth, nose \\
Modulation scheme   & Intensity-based modulation            & Respiratory-event-driven MoVCSK \\
Signal power        & Number of molecules                   & Number of particles \\
Events              & Clock synchronized                    & Random respiratory events \\
Information content & Data bits                             & Pathogens vs non-infected particles \\
Channel model       & Turbulent atmospheric channel         & Turbulent atmospheric channel \\
Network topology    & Point-to-point or point-to-multipoint & Point-to-multipoint; store and forward relaying \\
Receiver            & Absorbing or non-absorbing sensor     & Absorbing body parts (incl.~mouth, nose, and hands) \\
Efficiency measure  & Reliability information               & Viral load \\
Goal                & Maximization of mutual information    & Minimization of mutual information/infection rate \\
\bottomrule
\end{tabular}
\label{table:duality}
\vspace{-1.2em}
\end{table*}


\subsection{Aerosol Emission}
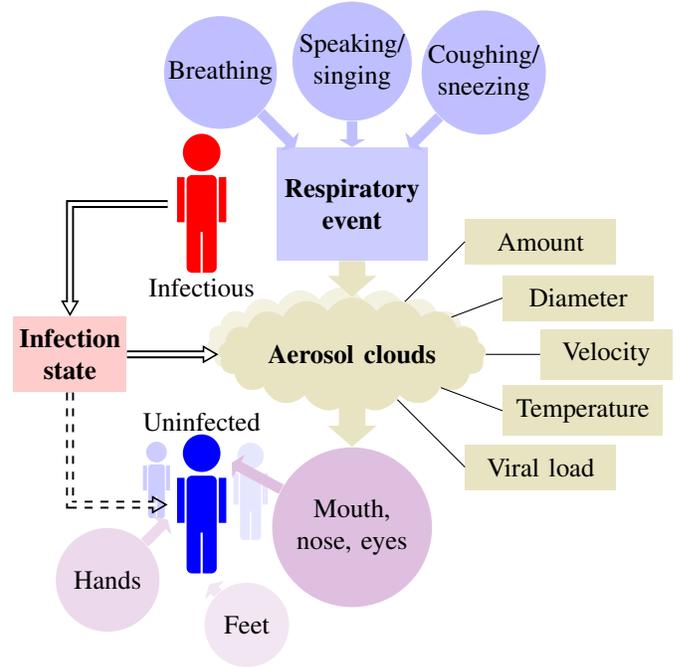
\begin{figure}
\addtolength\abovecaptionskip{-15pt}  
    \centering
    \input{2_emission_reception}
    \caption{Aerosol emission through a respiratory event and reception of aerosol clouds by other users. The aerosol distribution parameters (right-hand side) are influenced by the type of respiratory event (top part) and the infection state of the user (left-hand side). The uninfected users (bottom part) receive the pathogen-laden aerosols by their reception areas.}
    \vspace{-1.2em}
    \label{fig:aerosol_emission_reception}
\end{figure}
The emission of aerosols and droplets is triggered by respiratory events such as breathing, speaking, singing, coughing, or sneezing. These actions can be divided into three categories according to their type of occurrence, {\color{revised} which can be seen in the top part of Fig.~\ref{fig:aerosol_emission_reception}}:
\begin{enumerate}
    \item \textit{Breathing} can be modeled as continuous emission, following a breathing pattern, mainly influenced by the physical stress on the human body. Other factors like size of the body or age may also be considered.
    
    \item \textit{Speaking/singing} can be described as a two-state Markov process with the two states being either silence or singing/speaking. The model is defined by the transition probabilities between the two states and retention probabilities for staying in the current state. 
    
    \item \textit{Coughing/sneezing} can be emulated as a Poisson random process. The density parameters are influenced by the user's infection state, resulting in a higher probability for an infected user to cough or sneeze than for a healthy user.
    
\end{enumerate}
{\color{revised} Based on these events, aerosols are emitted into the environment, the main parameters of which are determined by the type of event. These parameters are}
\begin{itemize}
    \item the number of emitted particles,
    \item the diameter distribution,
    \item the initial velocity distribution,
    \item the temperature distribution, and
    \item the viral load or viral distribution.
\end{itemize}

Each respiratory event causes a time-varying particle cloud, even in the absence of air movement. Due to a counteraction between air drag, buoyancy, and gravity, aerosols survive for a longer period than droplets~\cite{Khalid2020,Gulec2020}. {\color{revised} The particle clouds of different respiratory events overlay and may react with each other. All parameters are of stochastic nature. The actual emitted aerosol distribution is user-dependent and may vary for each emission.  Larger droplets may shrink in dry air or split into smaller ones.}  The infection state of the human {\color{revised} influences the viral load and} ultimately decides if an emission contains pathogens or not, so every healthy user emits aerosols, but these have no influence on the infection status of the other users when received.

As an alternative to the human experiments conducted in~\cite{Schurwanz2020}, the mentioned respiratory events can be emulated in the air-based macro-scale testbed introduced in~\cite{Bhattacharjee2020a}. Mouth and nose are replaced by transmitter-side sprayers. In order to visualize particles, a fluorescent dye solution is emitted. Intensity and duty cycle are software-controlled. The angle of departure is adjustable. 
All mentioned actions can be emulated by respiratory-event-driven higher order {\color{revised} MoVCSK modulation~\cite{Schurwanz2020}.} 
In this context, an MoVCSK sender is capable to mimic infected and non-infected particles. This can be further refined by distinguishing between particles of different sizes. The intensities of the respiratory events are mapped to the different concentration levels. The random occurrence of the events over time can be represented by a sequence of MoVCSK symbols.

\subsection{Aerosol Reception}
Several parts of the human body are actively or passively involved in the absorption of aerosols, {\color{revised} as depicted in the bottom part of Fig.~\ref{fig:aerosol_emission_reception}}. The facial region with the mouth, the nose and the eyes is the most sensitive receiving area \cite{Gulec2020}, implying that a small amount of received aerosols can lead to an infection of the user. Other parts of the body like hands and feet or shoes are passively involved by receiving infected aerosols and either transferring them to more sensitive parts of the user by smear infection or contaminating other surfaces, which in turn can cause infection.

The reception-sensitive body areas can be interpreted as effective apertures, similar to spatially distributed antennas, with individual antenna gains and sensitivity levels. Starting with the highly sensitive body regions in the face, which can trigger an infection even at a low viral load, to the feet or shoes, which collect viruses settled at the floor. The sensitivity levels of the individual sensors can be modeled as user-dependent infection thresholds, which rely upon other criteria as well including the health conditions {\color{revised} and individual and cultural hygiene habits}.

The amount of viral molecules that the receiver absorbs per time unit, i.e., the viral load, has an impact on the probability of infection. According to the state-of-the-art, the human-dependent infection threshold is checked by hypothesis testing~\cite{Khalid2020, Gulec2020}: the probability of infection corresponds to the situation where the viral load exceeds the threshold. However, the viral load additionally is likely to have an influence on the severity of the course of the disease. This situation is related to reliability information. In communication and information theory, receiver-side reliability information describes the quality of information transmission in terms of the error rate.

From the perspective of our testbed~\cite{Bhattacharjee2020a}, the water-based dye solution from the sprayer is monitored and recorded by a digital camera at the receiver side under the influence of an ultraviolet (UV-A) light source. The intensity recorded by the camera can be interpreted as the viral load carried by the transmitted droplets over a certain distance through active transmission. The region between the sprayer and the camera contains those droplets which settle down under the influence of gravity. The intensities from these droplets can be recorded to formulate a model for air-based transmission of pathogens.

\subsection{Turbulent Atmospheric Channel}
The spatially-distributed time-varying aerosol clouds that are emitted into the environment from the various users, are subject to a dynamic channel with turbulences and changing parameters. Various effects change the channel and cause the aerosols to propagate through the space with different velocities and ranges. The turbulences in the air are caused by the movement of the users, ventilation devices, air streams introduced through windows and doors, temperature gradients, and weather phenomena. Turbulence can either increase the range of the aerosols by constructively influencing the propagation vector, or reduce it when the particles are forced to the ground. {\color{revised} Depending on the initial conditions in the environment, the range of the emitted aerosols can be influenced from the beginning.} In the testbed environment, turbulence can be introduced by introducing an external fan into the setup to cause an artificial air movement.


\subsection{Multiuser Scenario}

The scenario under investigation comprises multiple users in a predefined environment.
{\color{revised} The users are stationary or move along random trajectories while aerosols are continuously emitted due to the aforementioned respiratory events.}

Initially, at the physical layer level, the multiuser scenario can be described by a point-to-multipoint network, the nodes of which comprise of the entire set of users. The infectious user broadcasts pathogens. Some of these can be absorbed by one or more healthy users. Those users can be modeled as store and forward relays, which after some delay {\color{revised} (as detailed in Sections~\ref{sec:introduction} and~\ref{sec:scenario})} may broadcast even more pathogens. 
{\color{revised} Accordingly, the number of transmitters is randomly distributed.}
When the number of infectious users increases on average, the reproduction factor is said to be greater than one.

{\color{revised} In our testbed~\cite{Bhattacharjee2020a}, a point-to-multipoint scenario can be emulated by a network of pipelines with multiple camera receivers placed along their lengths with stationary or mobile sprayers releasing the water-based dye solution. If the intensities recorded by the cameras breach a certain threshold level, the receivers may be termed as ``infected'' actuating new sprayers after a certain delay, mimicking a relay mechanism.}

The users are able to dynamically move and change their absolute and relative positions. The absolute position affects the random emission pattern and causes spatially distributed aerosol clouds. Depending on the channel characteristics, the aerosols are able to stay suspended in the air for some time, resulting in a risk of infection for other users that move with their apertures through the aerosol clouds or that are hit by moving clouds.

Infection prevention by means of ``social distancing'' is focused on the relative positions among the users. Keeping appropriate relative distances may lead to a reduction in {\color{revised} infection count}, but ignores the possibility of the presence of suspended aerosols in the environment from previous respiratory events from other users at the same absolute position. 

In addition, the movement of the users create movement and turbulence of the surrounding air, which in itself results in a change of aerosol movement. 

%% file: 2_emission_reception.tex
\usetikzlibrary{calc}                   
\usetikzlibrary{shapes}      
\usetikzlibrary{arrows,arrows.meta,decorations.pathmorphing,backgrounds,positioning,fit,petri}
\begin{tikzpicture}
\tikzstyle{vecArrow} = [thick, decoration={markings,mark=at position
   1 with {\arrow[semithick]{open triangle 60}}},
   double distance=1.4pt, shorten >= 5.5pt,
   preaction = {decorate},
   postaction = {draw,line width=1.4pt, white,shorten >= 4.5pt}]
\tikzstyle{vecArrowDashed} = [thick,dashed,decoration={markings,mark=at position
    1 with {\arrow[semithick]{open triangle 60}}},
    double distance=1.4pt,shorten >= 5.5pt,
    preaction = {decorate},
    postaction = {draw,solid,line width=1.4pt,white,shorten >= 4.5pt}]
\tikzstyle{largerArrow} = [thick,decoration={markings,mark=at position
    1 with {\arrow[semithick]{open triangle 60}}},
    double distance=1.4pt,shorten >= 5.5pt,
    preaction = {decorate}]
\tikzstyle{innerWhite} = [semithick, white,line width=1.4pt, shorten >= 4.5pt]
\tikzstyle{betterArrow} = [-{Triangle[width=18pt,length=8pt]},line width=10pt]
\tikzstyle{betterArrowSmall} = [-{Triangle[width=10pt,length=3pt]},line width=4pt]

    \node (tx) at (0,0) {
    \begin{tikzpicture}
    \node[circle,fill=red,minimum size=5mm] (head) {};
    \node[rounded corners=2pt,minimum height=1.3cm,minimum width=0.4cm,fill=red,below = 1pt of head] (body) {};
    \draw[line width=1mm,red] ([shift={(2pt,-1pt)}]body.north east) --++(-90:6mm);
    \draw[line width=1mm,red] ([shift={(-2pt,-1pt)}]body.north west)--++(-90:6mm);
    \draw[thick,white] (body.south) --++(90:5.5mm);
    \end{tikzpicture}};
    \node [below=-0.2 of tx] (infected) {Infectious};
    
    \node[inner sep=0pt,minimum width=1.5cm,minimum height=1cm,color=black,fill=red!20,align=center,font=\bfseries] at (-1.75,-2) (infect) {Infection\\state}; 
    
    \node[rectangle,inner sep=0pt,minimum width=2cm,minimum height=1.5cm,color=black,fill=blue!20,align=center,font=\bfseries] at ($(tx)+(2,0)$) (event) {Respiratory\\event}; 
    
    \node[circle,inner sep=0pt,minimum size=1.5cm,color=black,fill=blue!25,align=center] at ($(event)+(-1.75,1.75)$) (breath) {Breathing}; 
    \node[circle,inner sep=0pt,minimum size=1.5cm,color=black,fill=blue!25,align=center] at ($(event)+(0,1.9)$) (speak) {Speaking/\\singing}; 
    \node[circle,inner sep=0pt,minimum size=1.5cm,color=black,fill=blue!25,align=center] at ($(event)+(1.75,1.75)$) (cough) {Coughing/\\sneezing}; 
    
    \node [cloud,cloud puffs=19,cloud puff arc=120, aspect=3,fill=olive!10,font=\bfseries] at ($(infect)+(+3.6,0.15)$) (cloud0) {Aerosol clouds};
    
    \node [cloud,cloud puffs=17,cloud puff arc=120, aspect=3,fill=olive!20,font=\bfseries] at ($(infect)+(+3.75,0)$) (aerosols) {Aerosol clouds};
    
    \node[inner sep=5pt,color=black,fill=olive!20,align=left,anchor=west,minimum height=0.6cm,minimum width=2cm] at ($(aerosols)+(1.5,1.5)$) (amount) {Amount}; 
    \node[inner sep=5pt,color=black,fill=olive!20,align=left,anchor=west,minimum height=0.6cm,minimum width=2cm] at ($(aerosols)+(2,0.75)$) (diam) {Diameter}; 
    \node[inner sep=5pt,color=black,fill=olive!20,align=left,anchor=west,minimum height=0.6cm,minimum width=1.75cm] at ($(aerosols)+(2.5,0)$) (velo) {Velocity}; 
    \node[inner sep=5pt,color=black,fill=olive!20,align=left,anchor=west,minimum height=0.6cm,minimum width=2cm] at ($(aerosols)+(2,-0.75)$) (temp) {Temperature}; 
    \node[inner sep=5pt,color=black,fill=olive!20,align=left,anchor=west,minimum height=0.6cm,minimum width=2cm] at ($(aerosols)+(1.5,-1.5)$) (load) {Viral load}; 
    
    \node (rx2) at (-0.6,-3.3) {
    \begin{tikzpicture}
     \begin{pgflowlevelscope}{\pgftransformscale{.55}}
    \node[circle,fill=blue!20,minimum size=5mm] (head) {};
    \node[rounded corners=2pt,minimum height=1.3cm,minimum width=0.4cm,fill=blue!20,below = 1pt of head] (body) {};
    \draw[line width=1mm,blue!20] ([shift={(2pt,-1pt)}]body.north east) --++(-90:6mm);
    \draw[line width=1mm,blue!20] ([shift={(-2pt,-1pt)}]body.north west)--++(-90:6mm);
    \draw[thick,white] (body.south) --++(90:5.5mm);
    \end{pgflowlevelscope}
    \end{tikzpicture}};
    
    \node (rx3) at (0.65,-3.35) {
    \begin{tikzpicture}
     \begin{pgflowlevelscope}{\pgftransformscale{.7}}
    \node[circle,fill=blue!10,minimum size=5mm] (head) {};
    \node[rounded corners=2pt,minimum height=1.3cm,minimum width=0.4cm,fill=blue!10,below = 1pt of head] (body) {};
    \draw[line width=1mm,blue!10] ([shift={(2pt,-1pt)}]body.north east) --++(-90:6mm);
    \draw[line width=1mm,blue!10] ([shift={(-2pt,-1pt)}]body.north west)--++(-90:6mm);
    \draw[thick,white] (body.south) --++(90:5.5mm);
    \end{pgflowlevelscope}
    \end{tikzpicture}};

    \node (rx) at (0,-4) {
    \begin{tikzpicture}
    \node[circle,fill=blue,minimum size=5mm] (head) {};
    \node[rounded corners=2pt,minimum height=1.3cm,minimum width=0.4cm,fill=blue,below = 1pt of head] (body) {};
    \draw[line width=1mm,blue] ([shift={(2pt,-1pt)}]body.north east) --++(-90:6mm);
    \draw[line width=1mm,blue] ([shift={(-2pt,-1pt)}]body.north west)--++(-90:6mm);
    \draw[thick,white] (body.south) --++(90:5.5mm);
    \end{tikzpicture}};
    \node [above=-0.2 of rx] (healthy) {Uninfected};
    
    \node[circle,inner sep=5pt,color=black,fill=violet!25,align=left,align=center] at ($(rx)+(2,-0.3)$) (head) {Mouth,\\nose, eyes}; 
    \node[circle,inner sep=5pt,color=black,fill=violet!15,align=left,align=center] at ($(rx)+(-1.25,-1)$) (hands) {Hands}; 
    \node[circle,inner sep=5pt,color=black,fill=violet!10,align=left,align=center] at ($(rx)+(0.6,-1.6)$) (feet) {Feet}; 
    
    \draw[vecArrow] (tx) -| (infect);
    \draw[vecArrowDashed] (infect) |- (rx);
    \draw[vecArrow] (infect) -- (aerosols);
    
    \draw[betterArrowSmall,blue!25] (breath) -- (event);
    \draw[betterArrowSmall,blue!25] (speak) -- (event);
    \draw[betterArrowSmall,blue!25] (cough) -- (event);
    
    \draw[betterArrow,olive!20] (event) -- (aerosols);
    \draw[betterArrow,olive!20] (aerosols) -- (head);
    
    \draw (aerosols) -- (amount.west);
    \draw (aerosols) -- (diam.west);
    \draw (aerosols) -- (velo.west);
    \draw (aerosols) -- (temp.west);
    \draw (aerosols) -- (load.west);

    \draw[betterArrowSmall,violet!25] (head) -- ($(rx.north)+(0.4,-0.5)$);
    \draw[betterArrowSmall,violet!15] (hands) -- ($(rx.west)+(0,-0.2)$);
    \draw[betterArrowSmall,violet!10] (feet) -- ($(rx.south)+(0.1,0)$);
\end{tikzpicture}

%% file: 3_minimization.tex
\section{Minimization of Mutual Information\label{sec:minimization_mutual_info}}
In this section, we apply the concept of mutual information to the area of infectious disease transmission. Contrary to a maximization of mutual information in data communication systems, the aim of infection prevention is to minimize the infection rate between nodes. This can be accomplished by employing several methods that are discussed afterwards. 

\subsection{Mutual Information in Pathogen-laden Transmission}

\begin{figure}
\addtolength\abovecaptionskip{-5pt}  
\addtolength\belowcaptionskip{-10pt}  
    \centering
    \input{3_z_channel}
    \caption{\color{revised} A point-to-point system can be simplified by the depicted discrete memoryless channel model, known as Z channel. Models like this serve as a basis for calculating the mutual information and the infection rate.}%
    \vspace{-1.2em}
    \label{fig:z_channel}
\end{figure}
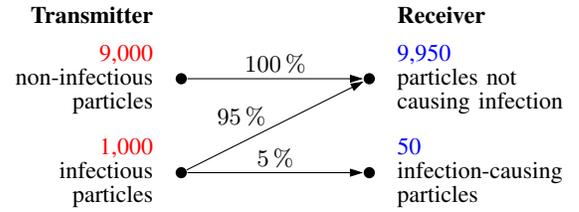

Infection is strongly related to the probability that a certain density of pathogen-laden particles successfully come into contact with the effective aperture area of the receiver. If the receiver-side density exceeds the user-dependent infection threshold, after a certain incubation time, the receiver will become infected with a certain probability.
In terms of Shannon information theory, this situation can be modeled mathematically by means of mutual information.

{\color{revised} The mutual information of a point-to-point communication channel is determined by the distribution of the channel input events as well as by the transition probabilities between channel input and channel output events~\cite{Akyildiz2019information}. In the scenario under investigation, the channel input distribution depends on the respiratory events. The transition probabilities depend on the statistics of the turbulent atmospheric channel, including receiver-side properties like the sensitive area~\cite{Hoeher2021}. An intuitive example is shown in Fig.~\ref{fig:z_channel}.  Out of 10,000 emitted particles, 10~\% are infectious.  This stipulates the input distribution.  Just 50 of the 1,000 emitted infectious particles hit the sensitive area of the receiver, corresponding to a transition probability of 5~\%. This concept can be generalized to multipoint-to-multipoint channel models. For details and a critical review of infection rate measures see \cite{Hoeher2021}.} The analogy between aerosol-based infection and mutual information demonstrates three important facts: the risk of infection depends on the channel input distribution including number of events, the turbulent channel, and the infection threshold. Although a minimization of mutual information is counterproductive in classical communication scenarios, it is a common design objective in crypto systems.

\subsection{Preventive Approaches to Reduce the Mutual Information}
\begin{figure}
\addtolength\abovecaptionskip{-5pt}  
    \centering
    \input{3_mutual_information_minimization.tikz}
    \caption{Possible mutual information minimization techniques to reduce the risk of infection.}
    \vspace{-1.3em}
    \label{fig:mutual_information_minimization}
\end{figure}
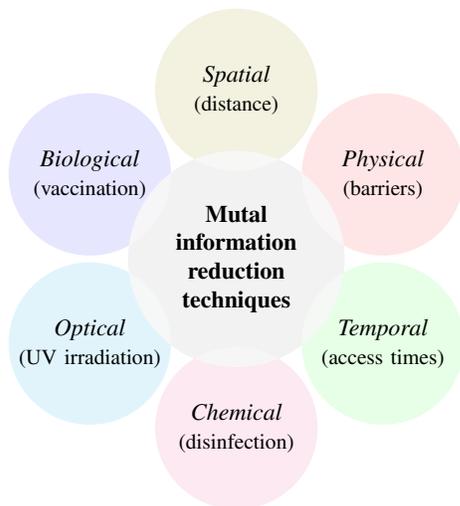
Fig.~\ref{fig:mutual_information_minimization} summarizes several techniques/approaches that reduce the risk of infection. These are described next in connection with mutual information. 

\textit{Spatial actions:} Maximization of Euclidean distances between the users in the multiuser scenario under investigation, i.e., ``social distancing'', aims to reduce the probability of direct infection caused by respiratory events. Other techniques like pruning/thinning the user density in a certain space as well as route optimizations to reduce the interference of users trajectories with infectious aerosol clouds have a similar impact on the propagation conditions.

Despite these spatial actions, sometimes it is desirable that the users are still able to communicate with each other. Hence, the goal is to enable communication to some extent while keeping the probability of infection at a tolerable level. This objective can be formulated as a constraint in an overall optimization problem. 

\textit{Physical actions:} Barriers and obstructions like face masks, face shields, and respirators or protective precautions like protective glasses or spit protections are widely applied techniques during the coronavirus pandemic. These actions have an impact on the channel input distribution, by reducing the amount of particles and the velocity of expelled air.

Another physical approach is air humidification. If the relative humidity of the room air is below 40 percent, the emitted particles absorb less water, stay lighter and move around the room for a longer period. In addition, the nasal mucous membranes in our noses become drier and more permeable to viruses when the air is dry. This has an impact on the infection threshold. Additional precautions like indoor fans, air ventilation, and air purification devices are also useful, as well as fresh air through open windows to prevent an accumulation of infectious aerosols in the ambient air.

\textit{Temporal actions:} The allocation of individual time slots on user or group level leads to a more orderly procedure and this results in a lower probability of infection. This technique is called time-division multiple access and is widely used in communication engineering. 
{\color{revised} Increasing the time between respiratory events reduces the likelihood of infection by pathogen-laden aerosols. An example is shift work.}

\textit{Chemical actions:} Reducing the viral load on surfaces is a key factor in infection prevention. This can be accomplished by cleaning of shoes or disinfection of the hands, as they are part of the effective aperture of the receiver. Household bleach solution containing 5.25 to 8.25 percent sodium hypochlorite could be used to disinfect surfaces while at least a 70 percent ethyl/isopropyl alcohol solution proves to be effective when it comes to proper disinfection of hands.  
{\color{revised} Disinfection and purifying sprays are a good alternative, as they can eliminate different kinds of viruses quickly. They might also stay suspended in air to further reduce the infectious particles over a longer period of time.}


\textit{Optical actions:} The viral load in airborne aerosols that were expelled into the air can be reduced by ultraviolet germicidal irradiation. This technique can be used in air purifiers to filter stale air, to clean handles, and other surfaces. It is essential to ensure that the skin and eyes are protected from radiation. Furthermore, UV-C irradiation can be employed for cleaning protection masks, even online when integrated into exhalation valves embedded in medical masks.  

\textit{Biological actions:} Medical methods with a direct effect on the receiver's detection mechanism provide the best protection against transmission of infection. At the same time, a biological solution is also the most difficult to apply to a wide range of users. An example of a biological action would be vaccination. At the same time, this is also the least broadly based mechanism, as vaccination usually only protects against a specific pathogen which is subject to a possible mutation. Ongoing research on biochemical nose sprays and mouthrinses is aimed at improving the infection threshold as well. 

Note that all these techniques/approaches have an influence on the mutual information. Some target the propagation conditions while others affect either the particle emission or the infection threshold, respectively.



%% file: 3_z_channel.tex
\begin{tikzpicture}
  \tikzstyle{every node}=[font=\small,execute at begin node=\setlength{\baselineskip}{1em}]
  
  \tikzstyle{betterArrow} = [-{Triangle[width=18pt,length=8pt]},line width=10pt]
  \tikzstyle{betterArrowSmall} = [-{Triangle[width=10pt,length=3pt]},line width=4pt]

  \node[align=right,anchor=east] at (-0.25,0.85) () {\textbf{Transmitter}};
  \node[align=left,anchor=west] at (2.75,0.85) () {\textbf{Receiver}};

  \node[align=right,anchor=east] at (-0.25,0) () {{\color{red}9,000}\\non-infectious\\particles};
  \node[align=right,anchor=east] at (-0.25,-1.25) () {{\color{red}1,000}\\infectious\\particles};

  \node[circle,fill,inner sep=1.5pt] at (0,0) (tx0) {};
  \node[circle,fill,inner sep=1.5pt] at (0,-1.25) (tx1) {};
  
  \node[circle,fill,inner sep=1.5pt] at (2.5,0) (rx0) {};
  \node[circle,fill,inner sep=1.5pt] at (2.5,-1.25) (rx1) {};
  
  \node[align=center] at (1.25,0.2) () {\SI{100}{\percent}};
  \node[align=center] at (0.8,-0.5) () {\SI{95}{\percent}};
  \node[align=center] at (1.25,-1.05) () {\SI{5}{\percent}};
  
  
  \node[align=left,anchor=west] at (2.75,0) () {{\color{blue}9,950}\\particles not\\causing infection};
  \node[align=left,anchor=west] at (2.75,-1.25) () {{\color{blue}50}\\infection-causing\\particles};
  
  \draw[-{Triangle[width=3pt,length=5pt]}] (tx0) -- (rx0);
  \draw[-{Triangle[width=3pt,length=5pt]}] (tx1) -- (rx0);
  \draw[-{Triangle[width=3pt,length=5pt]}] (tx1) -- (rx1);
\end{tikzpicture}

%% file: 3_mutual_information_minimization.tikz
\begin{tikzpicture}[scale=0.9, transform shape]
  \begin{scope}[blend group = soft light]
    \fill[gray!10] (0:0) circle (1.6);
    \fill[red!10] (30:2.5) circle (1.2);
    \fill[olive!10] (90:2.5) circle (1.2);
    \fill[blue!10] (150:2.5) circle (1.2);
    \fill[cyan!10] (210:2.5) circle (1.2);
    \fill[magenta!10] (270:2.5) circle (1.2);
    \fill[green!10] (330:2.5) circle (1.2);
  \end{scope}
  \node [align=center] at (30:2.5) {\textit{Physical}\\\small(barriers)};
  \node [align=center] at (90:2.5) {\textit{Spatial}\\\small(distance)};
  \node [align=center] at (150:2.5) {\textit{Biological}\\\small(vaccination)};
  \node [align=center] at (210:2.5) {\textit{Optical}\\\small(UV irradiation)};
  \node [align=center] at (270:2.5) {\textit{Chemical}\\\small(disinfection)};
  \node [align=center] at (330:2.5) {\textit{Temporal}\\\small(access times)};
  \node [align=center,font=\bfseries] {Mutal\\information\\reduction\\techniques};
\end{tikzpicture}

%% file: 4_simulation.tex

\section{Advanced Aerosol Simulation\label{sec:simulation}}
{\color{revised}
To gain further insight into such multiuser scenarios as described above, computer simulations can be used.
Compared to pure mathematical analysis, simulations afford flexibility in placing transmitters, receivers, and obstacles at arbitrary positions.
Simulations also offer expandability in keeping the door open for future implementations of additional physical phenomena such as Brownian motion or other mobility pattern.}

The propagation of aerosols can be simulated with high accuracy with the help of existing computational fluid dynamics (CFD) tools~\cite{abuhegazy2020numerical}.
This approach, however, is very demanding on computational resources.
If the goal is to simulate the spreading of disease in a multiuser scenario with mobility and over longer periods of simulated time, MC simulation tools optimized for the simulation of pathogen-laden particle transfer may become useful.
{\color{revised}
Using a state of the art particle simulator (e.g., \cite{Schurwanz2020}), however, comes with a trade-off in physical accuracy as, for example, air movement and turbulence due to ventilation or convection is treated in an abstract way.
In contrast, CFD helps to accurately assess the way in which infectious aerosols spread in an enclosed space (e.g., to investigate the effectiveness of ventilation).}

In previous work~\cite{Schurwanz2020}, we have extended the Pogona\footnote{\texttt{https://www2.tkn.tu-berlin.de/software/pogona/}} MC simulator presented in~\cite{drees2020efficient} aiming to replicate a coughing event.
In this paper, this scenario is extended further towards a multiuser network.
An emitter as presented in~\cite{Schurwanz2020} is an object which can be freely moved or duplicated in the simulation space.
At a specified time, the emitter releases particles in intervals for a fixed total duration.
The angle and velocity of emission are determined based on the configured probability distribution.
After emission, the trajectory of each particle is guided by its inertia and the forces acting on it, namely air drag, buoyancy, and gravity.
When a particle eventually collides with an absorbing receiver such as the ground or a spherical shape, it will be removed from the simulation and the position and speed of collision can be logged.

{\color{revised}\emph{Example:} Let us consider a scenario with three people.}
Each of the three people is modeled as having one point emitter at mouth height (\SI{1.65}{\meter}), one absorbing spherical receiver with a radius of \SI{5}{\centi\meter} immediately behind the point of emission, as well as two additional spherical receivers of the same size to model the hands at a height of \SI{70}{\centi\meter} with a lateral offset of \SI{30}{\centi\meter}.
The emitter releases 20 particles in intervals of \SI{0.25}{\milli\second} for a fixed duration of \SI{100}{\milli\second}.
Initially, person A represents an emitter and coughs with the parameters we determined experimentally in~\cite{Schurwanz2020}.
As one alteration to these parameters, the emission opening angle is now sampled randomly for each ejected particle based on the empirical cumulative distribution function of observed emission angles.
Previously, the emission opening angle was approximated with a normal distribution.
At a distance of \SI{2}{\meter} opposite of person A, person B is a passive receiver.
\SI{75}{\centi\meter} to the left of person B is person C, who acts as both a receiver as well as a second transmitter with a delay of \SI{3}{\second}.
In our previous work, the trajectory of simulated particles was only determined by their initial velocity vector at the time of emission and the forces acting on them, assuming a completely stationary volume of air for computing the drag force.
In order to emulate the initial ejection of air in a coughing event in a simplified fashion, we can emit the simulated particles inside a volume of constant air speed.
For this demonstration, we chose a \SI{3}{\meter} long and \SI{1}{\meter} wide cylindrical volume aligned with the vector of emission at the mouth, and an air speed of \SI{5}{\meter\per\second}. 

\begin{figure}
\addtolength\abovecaptionskip{-10pt}  
    \centering
    \includegraphics[width=\linewidth]{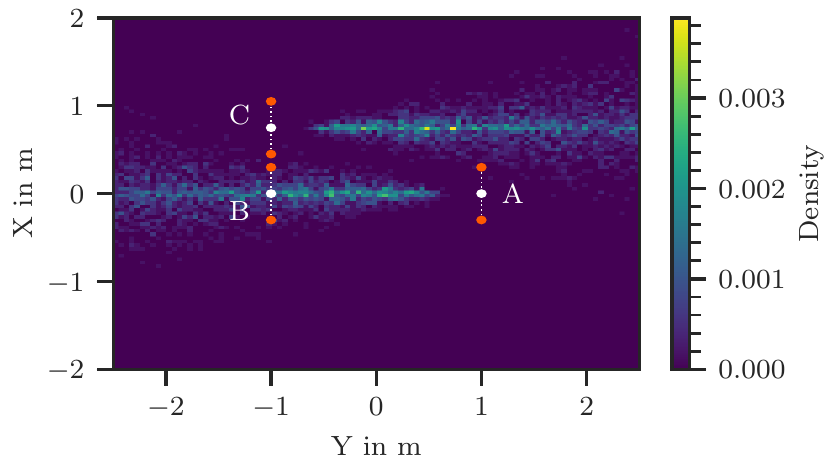}
    \caption{Density map of simulated particles intersecting with a horizontal plane at a height of \SI{0.65}{\meter} just below the hands. White circles represent the positions of the mouth regions and red circles represent the hands of the receivers.}%
    \vspace{-1.3em}
    \label{fig:simulation}
\end{figure}

As shown in Fig.~\ref{fig:simulation}, the receiving person B standing \SI{2}{\meter} in front of the emitter A is surrounded by the bulk of particles that have been emitted.
This receiver's facial region has received 112 simulated particles and the hands have received 14 particles, out of a total of 8000 emitted particles per cough.
The second receiver, offset by \SI{75}{\centi\meter} to the left of the first receiving person, did not receive any particles from the cough.
When this receiver C becomes the second emitter, the situation is mirrored and the first emitter A, now a receiver, is only reached by two particles at the closest hand.

{\color{revised} In future work, this simulation concept can be extended further to account for both the aspect of mobility and the remaining types of aerosol emissions outlined in Section~II. Mobility can be realized, for example, by letting simulated persons follow predetermined paths collected from mobility models or traces.  When incorporating other emission types, e.g., using the  air-based testbed outlined in Section~II, MoVCSK can be used as a modulation scheme to control the emission of particles.  In this context, it will also be important to support aerosol  particles with diameters smaller than \SI{50}{\micro\meter}.  Doing this accurately is difficult without  resorting to computation-intensive multiphysics and CFD simulations.  A sufficiently accurate  approximation could be obtained by expanding the original concept of the Pogona simulator \cite{drees2020efficient} and to guide aerosol particles through a sequence of samples of a time-varying pre-computed and locally constrained vector field.  Also, particle simulation tools could be used to support simulation tools on social (``networking'') layer.} 

%% file: 5_open_tasks.tex
\section{Prospects and Challenges\label{sec:open_questions}}
{\color{revised} Over time, much research has been done in the field of infection prevention. It is already possible to predict the course of infections in defined scenarios and to estimate the consequences of a certain number of infections. Techniques such as ``social distancing'' are results of this work.

The use of mutual information as a measure of the infection rate provides a common framework for different areas including engineering and infectiology. Emphasis is on a conceptual level by pointing out dualities and similarities between these fields in order to widen the view. New findings from existing and refined MC techniques can be applied in infectiology and evaluated from a health care point of view.  These may include improved distance rules for certain situations or guidance on placing novel sensors. Under the premise of enabling everyone to ``live with the virus'' with as few restrictions as possible, it is desirable to adapt the infection control solutions more precisely to the circumstances in specific situations.

Furthermore, a reverse check of established infection prevention techniques can be accomplished by means of tools established in communication and information theory. 
Nevertheless, especially when combining different disciplines like engineering and health care, it should be checked whether the proposed approaches are fundamentally correct, which in turn might lead to new prevention approaches being useful in the real world and thus reducing the incidence of infection.

Air-based MC offers new possibilities regarding the investigation of aerosol propagation. Testbeds and measurement tools can be used to verify precautions and to gain further insights into infection paths. Society would benefit from a better understanding of the spread of virus infections, not just Corona.
}

%% file: 6_conclusion.tex
\section{Conclusion\label{sec:conclusion}}
{\color{revised} This article extends recent advancements with respect to the symbiosis of air-based macro-scale MC and infectious particle transmission~\cite{Khalid2019, Khalid2020, Gulec2020, Schurwanz2020}. Emphasis is on a conceptual level by pointing out dualities and similarities between these fields.} In our concept, users are modeled as mobile nodes in a multiuser scenario. Infected users emit aerosol clouds in the sense of broadcasting, whereas healthy users may become infected with a certain probability if they absorb a sufficient viral load by viral-sensitive aperture areas, like spatially distributed antennas in the area of wireless communications. In turn, after some incubation time, these users may become pathogen-laden aerosol spreaders or superspreaders as well, related to store and forward relaying. Apart from being a pure binary problem of getting infected or not, the viral load frequently has an impact on the course of the disease, similar to reliability information in MC. In the sense of Shannon information theory, the goal is to minimize the mutual information between infected and non-infected users, where the information corresponds to infected and non-infected particles.

In this stochastic approach of infection prevention, numerous actions are discussed regarding channel input distribution, channel propagation, and channel output characteristic. For example, protective masks affect the channel input distribution. Spatial and temporal actions mainly have an impact on the channel propagation.
{\color{revised} Biochemical actions as well as health conditions affect the infection threshold at the channel output.} Additionally, challenges regarding {\color{revised} the objective to ``live with the virus''} are discussed. The conceptual approach is supported by an experimental macro-scale molecular communication testbed and by an advanced simulation tool. As an example, simulation results are presented for a 3D environment with three users who emit aerosols that are subject to air movement due to the ejection action. 

%% file: 7_appendix.tex

%



%% file: 8_acknowledgement.tex
\section*{Acknowledgment}
Reported research was supported in part by the project MAMOKO funded by the German Federal Ministry of Education and Research (BMBF) under grant numbers 16KIS0915 and 16KIS0917.